\documentclass[aps,prd,amsfonts,preprint,
eqsecnum,nofootinbib]
{revtex4}
\usepackage{graphics,epsfig}

\begin{document}
\preprint{WM-01-114}
%
\title{\vspace*{0.3in}
Discerning Noncommutative Extra Dimensions
\vskip 0.1in}
\author{Carl E. Carlson}
\email[]{carlson@physics.wm.edu}
\author{Christopher D. Carone}
\email[]{carone@physics.wm.edu}

\affiliation{Nuclear and Particle Theory Group, Department of
Physics, College of William and Mary, Williamsburg, VA 23187-8795}
\date{December, 2001}
\begin{abstract}
Experimental limits on the violation of four-dimensional Lorentz 
invariance imply that noncommutativity among ordinary spacetime 
dimensions must be small.  Noncommutativity among extra, 
compactified spatial dimensions, however, is far less constrained 
and may have discernible collider signatures.  Here we study the 
experimental consequences of noncommutative QED in six 
dimensions, with noncommutativity restricted to a TeV-scale 
bulk.   Assuming the orbifold $T^2/\mathbb Z_2$, we construct the 
effective four-dimensional theory and study interactions unique 
to the noncommutative case.  New vertices involving the 
Kaluza-Klein (KK) excitations of the photon yield order 100\% 
corrections to the pair production and to the decays of some of 
the lighter modes. We show that these effects are difficult to 
resolve at the LHC, but are likely within the reach of a future 
Very Large Hadron Collider (VLHC).
\end{abstract}
\maketitle


\section{Introduction}\label{sec:intro}


Interest in the possibility of extra, compactified spatial 
dimensions at the TeV scale~\cite{led} has led to serious consideration 
of other modifications of  spacetime structure that may be manifest 
at collider energies.  One such possibility is that ordinary four-dimensional 
spacetime may become 
noncommutative~\cite{HPR,HK,GD,Mah,BGXW,GL,ACDNS,MPR,CHK,ABDG,CST1,CCL,RJ} 
at some (not so) high scale $\Lambda_{{\rm NC}}$: promoting the position 
four-vector $x^\mu$ to an operator $\hat x^\mu$, one assumes that
\begin{equation}\label{eq:com}
\left[\hat{x}^\mu \, , \, \hat{x}^\nu \right]= i \theta^{\mu\nu}
\end{equation}
where $\theta$ is a real, constant matrix with elements of order 
$(\Lambda_{{\rm NC}})^{-2}$.  Field theories defined on such a 
noncommutative spacetime involve field operators that are functions of the 
noncommuting coordinates $\hat{x}^\mu$.  However, it is possible to map 
such a theory to a physically equivalent one involving fields that are 
functions of the classical coordinates $x^\mu$.  Given a classical 
function $f(x)$ with Fourier transform
\begin{equation}
\tilde{f}(k)=\frac{1}{(2\pi)^{n/2}} \int d^n x e^{ik_\mu x^\mu} f(x) \,\,\, ,
\end{equation}
one may associate the operator
\begin{equation}
W(f)=\frac{1}{(2\pi)^{n/2}} \int d^n k e^{-ik_\mu \hat{x}^\mu} \tilde{f}(k)
\end{equation}
in the noncommuting theory~\cite{Madore}.  Requiring that this correspondence 
holds for the product of functions,
\begin{equation}
W(f) W(g) = W(f \star g)
\end{equation}
one finds that
\begin{equation}\label{eq:mstar}
f \star g = \lim_{y\rightarrow x} e^{\frac{i}{2} \frac{\partial}
{\partial x^\mu}
\theta^{\mu\nu}\frac{\partial}{\partial y^\nu}} f(x) g(y)  \,\,\, .
\end{equation}
This is the Moyal-Weyl $\star$-product~\cite{mwsp}. The starting point 
for most of the phenomenological study of noncommutative field theories is 
the construction of gauge-invariant Lagrangians in which ordinary 
multiplication has been replaced by the star product.  The resulting 
classical field theory is amenable to quantization and study in the 
usual ways.

Noncommutative field theories (NCFTs) have been shown to arise in the 
low-energy  limit of some string theories~\cite{Seiberg}, and this has 
partly motivated a number of the recent phenomenological 
studies. On the other hand, NCFTs have been studied frequently without 
reference to higher-energy embeddings, and are interesting as effective 
field theories by themselves~\cite{Madore,HAY,ARM,LIA,Jurco,CSJT,CJSWW}.  
We will view the noncommutative theory that we study as a low-energy 
effective theory, valid up to the Planck scale, which may be relatively low.  

Prior phenomenological work on noncommutative extensions of the standard 
model have focused on direct collider searches~\cite{HPR,HK,GD,Mah,BGXW,GL}, 
astrophysical effects~\cite{ACDNS}, and indirect 
low-energy bounds~\cite{MPR,CHK,CST1,ABDG,CCL}.  The papers on collider 
signatures have dealt exclusively with noncommutative QED (NCQED), 
primarily for two reasons.  First, up until recently~\cite{Jurco,CSJT,CJSWW}, 
it was only known how to formulate noncommutative U(1)~\cite{HAY}
and U(N)~\cite{ARM} gauge theories consistently, with the matter content 
of the former restricted to charges of $0$ or $\pm 1$.  Secondly, NCQED 
has interaction vertices that are strikingly different from those in 
ordinary QED. In particular, the {\em Abelian} field strength tensor in 
NCQED has the form
\begin{equation}  
F_{\mu\nu}=\partial_\mu A_\nu-\partial_\mu A_\nu-i A_\mu \star A_\nu+
i A_\nu \star A_\mu  \,\,\, .
\end{equation}
Notice that the last two terms cancel in the $\theta \rightarrow 0$ limit, 
but otherwise lead to exotic three- and four-point photon vertices.  
Provided that the scale $\Lambda_{{\rm NC}}$ is sufficiently low, such 
interactions could be discerned, for example, at the NLC~\cite{HPR,GD,Mah,
BGXW,GL}.

The prior work on low-energy tests of 
noncommutativity~\cite{MPR,CHK,CST1,ABDG,CCL} has focused mainly on
searches for Lorentz violation~\cite{lv}, arising from the 
constant parameter $\theta^{\mu\nu}$ appearing in Eq.~(\ref{eq:com}).  
Noting that $\theta^{\mu\nu}$ is antisymmetric, one can regard $\theta^{0i}$ 
and $\theta^{ij}$ as constant three-vectors indicating preferred directions in 
a given Lorentz frame.  Low-energy tests of Lorentz invariance place bounds 
on $\Lambda_{{\rm NC}}$ of order $10$~TeV~\cite{CHK}, if one considers 
NCQED processes at tree-level.  As was pointed out by Mocioiu 
{\em et al.}~\cite{MPR}, the generation 
of an effective operator of the form ${\bar N} \theta^{\mu\nu} 
\sigma_{\mu\nu}N$ leads to a shift in nuclear magnetic moments and an 
observable sidereal variation in the magnitude of hyperfine splitting in 
atoms. The resulting bound on $\theta^{-1/2}$ is of the same order as the 
conventional (nonsupersymmetric) grand unification scale.  That similar
operators are generated radiatively at the quark level was demonstrated 
in a number of toy models by  Anisimov {\em et al.}~\cite{ABDG}, and in 
a consistent formulation of noncommutative QCD by 
Carlson {\em et al.}~\cite{CCL}.  In the latter work, for example, the 
authors obtained the restriction $\theta \Lambda^2 < 10^{-29}$, where 
$\theta$ is a typical entry in $\theta^{\mu\nu}$, and $\Lambda$ is an 
ultraviolet regularization scale.  For 
$\Lambda \sim M_{{\rm Planck}} \sim 1$~TeV, the Lorentz-violation from 
such radiatively-generated, lower-dimension operators seems to imply that 
the size of $\theta$ must be much smaller than one might estimate based 
on any naturalness arguments.  

If one accepts the generic conclusions of Refs.~\cite{MPR,ABDG,CHK,CCL}, 
then colliders have little chance of probing noncommutative 
phenomenology.  In this paper, we point out that there is a way 
around this conclusion in the case where noncommutativity is 
restricted to extra spatial dimensions.  For concreteness, we 
will consider NCQED in six dimensions, where two extra spatial 
dimensions are compactified on the toroidal orbifold  $T^2/\mathbb Z_2$.  
The compact, extra-dimensional space $T^2/\mathbb Z_2$ is the minimal
choice that can be noncommutative and yield a phenomenologically viable
low-energy theory.  Orbifolding the torus allows one to project out
unwanted scalar photon zero modes in the effective 4D theory and
$\mathbb Z_2$ is the smallest discrete symmetry that accomplishes the 
task. (For other scenarios see, for example, Refs.~\cite{MPR2,GMW}.)
Since ordinary four-dimensional spacetime is commutative, with 
itself and with the extra dimensions, there is no violation of 
four-dimensional Lorentz invariance, and the most stringent 
bounds described above are evaded. Higher-dimensional Lorentz 
invariance is broken through compactification in any case, so 
that the phenomenological constraints should be no 
stronger than in the case of commutative extra dimensions.  We 
will show that new, $\theta$-dependent interactions are present 
in our theory, but involve exclusively the Kaluza-Klein (KK) 
excitation of the photon. We consider how these interactions may 
be discerned at hadron colliders through the decays and through pair 
production of some of the lighter modes.


\section{Formalism}


We consider six-dimensional (6D) quantum electrodynamics
(QED) with the gauge fields defined on the full space and
the fermion fields restricted to a 4D subspace.  The
Lagrangian is
\begin{eqnarray}
{\cal L}_6 &=& - {1\over 4} {\cal F}_{MN} \star {\cal F}^{MN} 
         + {\cal L}_{gauge\ fixing}
                              \nonumber \\ 
        &+& \delta^{(2)}(\vec y) \left\{
           \bar\psi (i\not\!\partial -m) \psi 
         + \hat e \bar\psi \star \not\!\!\! {\cal A} \star \psi
                                          \right\}  \ ,
\end{eqnarray}

\noindent where 
\begin{equation}
{\cal F}_{MN} = \partial_M {\cal A}_N - \partial_N {\cal A}_M 
               -i \hat e \left[{\cal A}_M 
                 \stackrel{\star}{,} {\cal A}_N
                                                \right]  \ ,
\end{equation}

\noindent and where $\hat{e}$ is the 6D gauge coupling.  The star ($\star$) 
indicates the Moyal product, Eq.~(\ref{eq:mstar}),
and the star commutator $[.\stackrel{\star}{,}.]$ indicates
a commutator with Moyal multiplication.  Our notation for the position
six-vector is 
\begin{equation}
X^M  = (x^0,x^1,x^2,x^3,y^5,y^6)  \ ,
\end{equation}

\noindent with $\vec y \equiv (y^5,y^6)$.  

\begin{figure}[ht]
\centerline {\epsfxsize 3. in \epsfbox{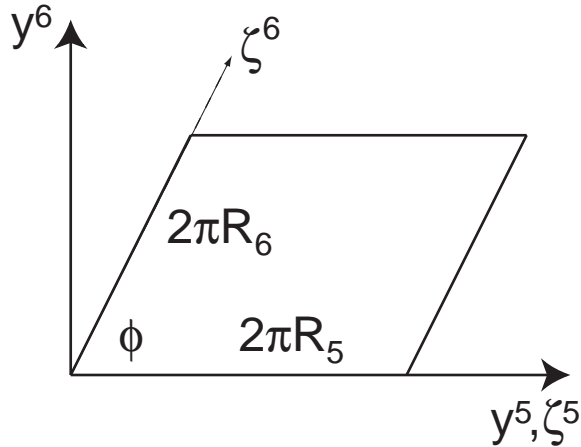}   }
\caption{The two dimensional torus with differing radii and shift angle
$\phi$.  Orthogonal axes are $y_i$ and skewed axes are $\zeta_i$.}
\label{skewedtorus}
\end{figure}

We compactify the extra dimensions on the orbifold 
$T^2/ \mathbb Z_2$, where $T^2$ is a general 2-torus.
We take into account the possibility of two different radii $R_5$ and $R_6$ 
and a relative angle $\phi$ between the two directions of
compactification~\cite{ant,kd}, as illustrated in Fig.~\ref{skewedtorus}. The
coordinates along the torus are $\zeta^i$, related to $y^i$ by
\begin{eqnarray}
y^5 &=& \zeta^5 + \zeta^6 \cos\phi  \nonumber \\
y^6 &=& \zeta^6 \sin\phi           \ .
\end{eqnarray}

\noindent The periodicity requirements on a function of orbifold 
coordinates $f(\zeta^5,\zeta^6)$ are
\begin{equation}\label{eq:reqs}
f(\zeta^5,\zeta^6) = f(\zeta^5 + 2\pi R_5, \zeta^6) 
= f(\zeta^5, \zeta^6 + 2\pi R_6)    \ .
\end{equation}

Without orbifolding, Eq.~(\ref{eq:reqs}) implies that bulk fields have
6D wave functions proportional to
\begin{eqnarray}
&&\exp \left\{ i {n^5 \zeta^5\over R_5 } +i {n^6 \zeta^6 \over R_6 }
\right\}
               \nonumber \\
&=& \exp  \left\{ i {n^5 y^5 \over R_5} + i {y^6 \over \sin\phi}
             \left[ {n^6\over R_6} - {n^5\over R_5}\cos\phi \right]
                                                \right\} \ ,
\end{eqnarray}

\noindent  where $n^5$ and $n^6$ are integers. The masses of the KK modes
are eigenvalues of the mass operator 
$-\partial^2_{y^5}-\partial^2_{y^6}$ and are given by
\begin{equation}
m_{\vec n}^2 = {1\over \sin^2 \phi} \left( {n_5^2 \over R_5^2}
                                        + {n_6^2 \over R_6^2}
                  - {2 n_5 n_6\over R_5 R_6} \cos\phi   \right)  \ ,
\end{equation}

\noindent where $\vec n \equiv (n^5,n^6)$.

\begin{figure}[ht]
\centerline {\epsfxsize 1.5 in \epsfbox{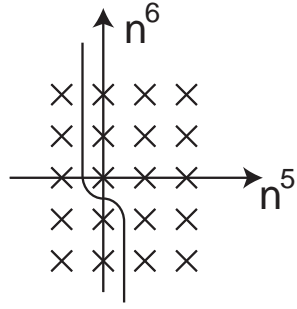}   }
\caption{The set $\{\vec n_+\}$ consists of the points to the right of
the line with the offset.}
\label{nc_art}
\end{figure}

The $\mathbb Z_2$ orbifolding consists of identifying points connected
by $\vec{y} \rightarrow -\vec{y}$~\cite{ant}.  Different
components of the gauge field may be Fourier expanded with 
different $\mathbb Z_2$-parities so that zero modes are only present
for the first four components,
\begin{eqnarray}
{\cal A}_M(X) = \sum_{\{\vec n_+\}} {\cal A}_M^{(\vec n)} (x) \left\{
  \begin{array}{ll}
    \cos \left( n^5 \zeta^5 + \xi n^6 \zeta^6 \over R \right)\, ,  & 
\qquad M = \mu  
                            \\[2ex]
    \sin \left( n^5 \zeta^5 + \xi n^6 \zeta^6 \over R \right)\, ,  & 
\qquad M = 5,6
  \end{array}   \right.
\end{eqnarray}

\noindent with $\mu=0,1,2,3$. Here, $\xi$ is the ratio of the radii,
with
\begin{equation}
R_5 \equiv R = \xi R_6   \ .
\end{equation}
Since orbifolding has provided wave functions with distinct parities, the
value of $\vec{n}$ is now restricted to a half plane
including the origin,
\begin{eqnarray}
{\{\vec n_+\}} = \left\{
  \begin{array}{l}
     \vec n = \vec 0; \quad {\rm or} \\
     n^5 = 0, n^6 > 0; \quad {\rm or} \\
     n^5 > 0, n^6 = {\rm any\ integer}
  \end{array}                              \right\} \ ,
\end{eqnarray}

\noindent also shown in Fig.~\ref{nc_art}.  Within the set
$\{\vec n_+\}$, masses are unique for most values of
$\xi$ and $\phi$.

One obtains the 4D Lagrangian by integrating over the extra dimensions,
\begin{equation}
{\cal L}_4 = \int d^2 y \, {\cal L}_6  
           = {1\over\xi_0} \int_0^{2\pi R} d\zeta^5 \, 
             \int_0^{2\pi R} d\tilde{\zeta^6} \,   {\cal L}_6 \ ,
\end{equation}

\noindent where $\xi_0 \equiv \xi / \sin\phi$ and 
$\tilde{\zeta^6}\equiv \xi \zeta^6$.

The gauge fixing Lagrangian is chosen as
\begin{equation}
{\cal L}_{gauge\ fixing} = -{1\over 2} \eta
        \left( \partial_\mu {\cal A}^\mu 
              + {1\over \eta} \partial_k {\cal A}^k  \right)^2
\end{equation}

\noindent with $k$ = 5,6. Terms in the Lagrangian quadratic in the
gauge field become,
\begin{eqnarray}
&&\left(  -{1\over 4} {\cal F}_{MN} {\cal F}^{MN} 
       + {\cal L}_{gauge\ fixing}  \right)_{free,\ 4d} \nonumber \\[2ex]
   &=& -{1\over 4} F_{\mu\nu}^{(\vec 0)} F^{\mu\nu (\vec 0)} 
     -{1\over 2} \eta \left( \partial^\mu A_\mu^{(\vec 0)} \right)^2
                                               \nonumber  \\
&+&{\sum}' \Bigg\{ -{1\over 4}  F_{\mu\nu}^{(\vec n)} F^{\mu\nu (\vec n)}
        -{1\over 2} \eta \left( \partial^\mu A_\mu^{(\vec n)} \right)^2
                                                          \\ 
&&  \qquad\ 
      +{1\over 2} \partial_\mu A_L^{(\vec n)}\partial^\mu A_L^{(\vec n)}
      +{1\over 2} \partial_\mu A_H^{(\vec n)}\partial^\mu A_H^{(\vec n)}
                                                \nonumber  \\ 
&+&     {1\over 2} m_{\vec n}^2    A^{\mu (\vec n)}   A_\mu^{(\vec n)}
   - {1\over 2} m_{\vec n}^2    A_L^{(\vec n)}   A_L^{(\vec n)}
   - {1 \over 2\eta} m_{\vec n}^2  A_H^{(\vec n)}   A_H^{(\vec n)}
                             \Bigg\}.   \nonumber
\end{eqnarray}

\noindent The primed sum is over the Kaluza-Klein (KK) modes, i.e., over
$\{\vec n_+\}$ excluding $\vec 0$.  The 6D fields $\cal A$ have been
rescaled, 
\begin{eqnarray}
{\cal A}_M^{(\vec 0)} &=&  { \sqrt{\xi_0} \over 2\pi R} A_M^{(\vec 0)} \ ,
                                        \nonumber \\
{\cal A}_M^{(\vec n)} &=&  { \sqrt{2\xi_0} \over 2\pi R} A_M^{(\vec n)}
                         \qquad  [\vec n \not= \vec 0] \ ,
\end{eqnarray}

\noindent where the fields $A$ have their canonical 4D mass dimensions. 
The fifth and sixth components have been combined into
\begin{eqnarray}
A_L^{(\vec n)} = {1\over |\vec {\tilde n}|}
           \left( \tilde n^5 A^{6 (\vec n)} - \tilde n^6 A^{5 (\vec n)}
                                  \right)  \ ,
                                            \nonumber \\
A_H^{(\vec n)} = {1\over |\vec {\tilde n}|}
           \left( \tilde n^5 A^{5 (\vec n)} + \tilde n^6 A^{6 (\vec n)}
                                            \right)  \ ,
\end{eqnarray}

\noindent where $\vec {\tilde n} =(\tilde n^5, \tilde n^6)$ with
\begin{eqnarray} \label{eq:ntilde}
\tilde n^5 &=& n^5  \nonumber \\
\tilde n^6 &=& {1\over \sin\phi}
            \left( \xi n^6 - n^5 \cos\phi \right)  \ ,
\end{eqnarray}

\noindent  and $m_{\vec n} = |\vec{\tilde n}|/R$.  The fields
$A_L$ and $A_H$ are physical and unphysical scalars in the 4D
theory, respectively. As $\eta \rightarrow 0$, the field $A_H$ is removed 
from the theory, the extra-dimensional generalization of unitary gauge.
We work in the $\eta \rightarrow 0$ limit henceforth.  Thus, from the 
free gauge Lagrangian the physical states are the
ordinary massless photon, the vector KK modes, and the scalar KK modes
$A_L^{(\vec n)}$.  

The fermion fields $\psi$ are defined only at the $\vec{y}=\vec{0}$ orbifold
fixed point, and involve no rescaling.  Since $A_L^{(\vec{n})}$ is odd under 
the $\mathbb Z_2$ parity it vanishes at $\vec{y}=\vec{0}$.
Hence the fermions interact only with the photon and its vector KK
excitations.  The fermion Lagrangian is 
\begin{eqnarray}
{\cal L}_{f,4d} = \bar\psi (i\not\! \partial &-& m) \psi
                + e \bar\psi \, \star \not\!\! A^{(0)} \star \psi
                                          \nonumber \\
                &+& e \sqrt{2} \, {\sum}' \, 
                  \bar\psi \, \star \not\!\! A^{(\vec n)} \star \psi  \ .
\end{eqnarray}

\noindent The 4D gauge coupling has been identified through the rescaling
\begin{equation}
\hat e = {2\pi R \over \sqrt{\xi_0}} e \ .
\end{equation}

\begin{figure}[ht]
\centerline {\epsfxsize 2 in \epsfbox{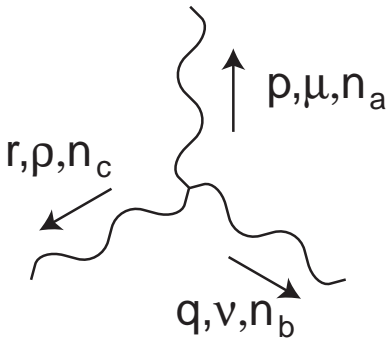}   }
\caption{The triple KK photon vertex.}
\label{feynman}
\end{figure}

The pure gauge field interactions come from the terms
\begin{eqnarray}
{\cal L}_{g\ int,6d} &=&  i \hat e  \partial_M {\cal A}_N 
         \left[{\cal A}^M  \stackrel{\star}{,} {\cal A}^N  \right]
                                  \nonumber \\
  &+& {1\over 4} \hat e^2
         \big[{\cal A}_M  \stackrel{\star}{,} {\cal A}_N  \big] \star
         \big[{\cal A}^M  \stackrel{\star}{,} {\cal A}^N  \big]  \ ,
\end{eqnarray}

\noindent in which the Moyal commutator may be written as
\begin{equation}
\left[{\cal A}_M  \stackrel{\star}{,} {\cal A}_N  \right]
  = 2i \lim_{X \rightarrow Y}  
           \sin\left( {1\over 2} 
   \frac{\partial}{\partial X^i} \theta^{ij} \frac{\partial}{\partial Y^j}
                  \right){\cal A}_M(X) {\cal A}_N(Y)  \ .
\end{equation}
One may now extract the three-photon coupling in the 4D Lagrangian,
\begin{eqnarray}
{\cal L}_{3\gamma,4d} &=& -e\sqrt{2} \, {\sum}' 
       \left( \delta_{\vec n_a, \vec n_b + \vec n_c}
          +   \delta_{\vec n_b, \vec n_c + \vec n_a}
          -   \delta_{\vec n_c, \vec n_a + \vec n_b}  \right)
                                      \nonumber \\
  &\times& \partial_\alpha A_\beta^{(\vec n_c)}
           A^{\alpha (\vec n_a)} A^{\beta (\vec n_b)}
  \sin\left(\tilde n_a^i \theta_{ij} \tilde n_b^j \over 2R^2 \right) \ ,
\end{eqnarray}

\noindent where $\tilde{n}$ is defined in Eq.~(\ref{eq:ntilde}).
The triple-photon couplings involve only the KK modes, and never any
ordinary massless photons.  The Feynman rule that corresponds to the
$3\gamma$ term in the Lagrangian, for the momenta, Lorentz indices, and
KK modes labeled in Fig.~\ref{feynman}, is given by
\begin{eqnarray}\label{eq:thevertex}
V_{3\gamma} &=& -e\sqrt{2} 
       \left( \delta_{\vec n_a, \vec n_b + \vec n_c}
          +   \delta_{\vec n_b, \vec n_c + \vec n_a}
          -   \delta_{\vec n_c, \vec n_a + \vec n_b}  \right)
                                      \nonumber \\
                        &\times&
   \sin\left(\tilde n_a^i \theta_{ij} \tilde n_b^j \over 2R^2 \right)
                                                \\
                        &\times&
  \left[ g_{\mu\nu} (p-q)_\rho +g_{\nu\rho} (q-r)_\mu
         +  g_{\rho\mu} (r-p)_\nu    \right]    \ .  \nonumber
\end{eqnarray}

When the noncommutativity is only in the
extra dimensions, the only independent non-zero component of the
noncommutativity tensor is $\theta^{56} \equiv \theta$.  
(Theories with space-like noncommutativity are known to preserve
perturbative unitarity~\cite{gomis}.) The argument of
the sine simplifies using
\begin{equation}
\tilde n_a^i \theta_{ij} \tilde n_b^j = \xi_0 \theta 
              \left(n_a^5 n_b^6 - n_a^6 n_b^5 \right) \ .
\end{equation}

A four-photon vertex may be computed in a similar way, but will
not be relevant to the physical processes studied in the sections
that follow.


\section{Decays} \label{sec:decays}

\begin{figure}[ht]
\centerline {\epsfxsize 3 in \epsfbox{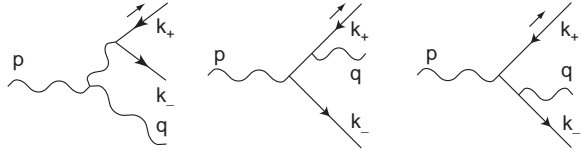}}
\caption{Feynman diagrams for the three-body decay $\gamma^{(\vec{n}_a)} 
\rightarrow \gamma^{(\vec{n}_b)} \, f \, \bar{f}$. including the
noncommutative triple KK photon vertex.}
\label{fig:dec}
\end{figure}

We first investigate the possibility of detecting noncommutativity in 
extra dimensions via corrections to the decays of the lighter KK modes.
We will discuss production rates of single KK modes later in this section.
The two-body decay $\gamma^{(\vec{n})}\rightarrow f {\bar f}$ is unaffected
by noncommutativity, while $\gamma^{(\vec{n}_a)}\rightarrow\gamma^{(\vec{n}_b)}
\,\gamma^{(\vec{n}_c)}$ is kinematically inaccessible for decays allowed
by Eq.~(\ref{eq:thevertex}).  However,
by taking one of the external lines in Fig.~\ref{feynman} off-shell, the new 
noncommutative vertex contributes at tree-level to the three-body decay 
$\gamma^{(\vec{n}_a)} \rightarrow \gamma^{(\vec{n}_b)} \, f \, \bar{f}$, 
where $f$ is a fermion at the $\vec{y}=0$ fixed point.  As one can see from 
Fig.~\ref{fig:dec}, the new contribution to the amplitude occurs at 
order $\alpha$, and is potentially as large as the `standard' diagrams.   
Henceforth, we use the term `standard' or `nonstandard' to refer to 
diagrams that are nonvanishing or vanishing in the $\theta \rightarrow 0$ 
limit.

For concreteness, let us choose the initial KK mode to be $\vec{n}_i=(1,1)$ 
and the on-shell KK mode in the final state to be $\vec{n}_f=(1,0)$.  We 
write the decay width
\begin{equation}
\Gamma = \Gamma_S + \Gamma_{NS}
\end{equation}
to distinguish the standard and nonstandard contributions.  
There is no interference between the standard and noncommutative diagrams 
because they are 90 degrees out of phase. We present our
results in terms of the differential decay width, written as a function of 
the energies of the outgoing fermions.  For the nonstandard diagram, we
find
\begin{equation}\label{eq:nsdecay}
\frac{d\Gamma_{NS}}{dE_+ dE_-} = \frac{16}{3}\frac{\alpha^2}{\pi}
\frac{1}{m_{11}} \sin^2 \left(\frac{\xi_0 \theta}{2 R^2}\right)
\cdot f(E_+,E_-) \cdot g(E_+,E_-)
\end{equation} 
where
\[
f(E_+,E_-)= \frac{(m^2_{21}-m^2_{01})^2}{
[2m_{11}(E_++E_-)+m^2_{10}-m^2_{11}-m^2_{01}]^2
[2m_{11}(E_++E_-)+m^2_{10}-m^2_{11}-m^2_{21}]^2} 
\]
\begin{equation}\end{equation}
and 
\[
g(E_+,E_-)=\frac{1}{2 m^2_{10}}[
8 E_-^2 E_+^2 m_{11}^2+4 (E_- E_+^3+E_+ E_-^3) m_{11}^2
+2 (E_+^3+E_-^3) (m_{11}^3+2 m_{10}^2 m_{11})
\]\[
-2(E_+ E_-^2+E_- E_+^2 )(m_{11}^3-6  m_{10}^2 m_{11})
+(E_+^2+ E_-^2 ) (-5 m_{11}^4-9 m_{10}^2 m_{11}^2+2 m_{10}^4) 
+2 E_+ E_- (-3 m_{11}^4
\]\[
-5 m_{10}^2 m_{11}^2+2 m_{10}^4)
+2 (E_+ +E_-) (2 m_{11}^5-m_{10}^2 m_{11}^3-4 m_{10}^4 m_{11})
+(-m_{11}^6+3 m_{10}^2 m_{11}^4-2 m_{10}^6)]
\]
\begin{equation}\end{equation}
For notational convenience, we have labelled the mass of the 
$\vec{n}=(i,j)$ mode as $m_{ij}$.  The function $f(E_+,E_-)$ originates 
from propagators of the off-shell KK modes; the dependence on
$m_{01}$ and $m_{21}$ reflects that only the $(0,1)$ and $(2,1)$ modes 
may contribute to the internal line, given the choice of
external states and the delta functions appearing in Eq.~(\ref{eq:thevertex}).
For the standard contributions, we find
\[
\frac{d\Gamma_{S}}{dE_+ dE_-} = \frac{8}{3}\frac{\alpha^2}{\pi}
\frac{1}{m_{11}} \left[ \frac{(m_{11}-2 E_+)(m_{11}-2 E_-)-m_{10}^2}
{(m_{11}-2 E_-)^2} \right.
\]
\[
+\frac{{(m_{11}-2 E_+)(m_{11}-2 E_-)-m_{10}^2}}{(m_{11}-2 E_+)^2}+
\]
\begin{equation}\label{eq:sdecay}
\left.
\frac{2 (m_{10}^2-m_{11}^2+2 (E_+ + E_-)m_{11}) 
(m_{11}^2+m_{10}^2)}{m_{11}^2 (m_{11}-2 E_-) 
(m_{11}-2 E_+)}\right]
\end{equation}
We present two quantities for our numerical results.  We evaluate
the partial decay width by integrating Eqs~(\ref{eq:nsdecay}) and
(\ref{eq:sdecay}) over the ranges
\begin{eqnarray}
-E_- + \frac{1}{2m_{11}}(m_{11}^2-m_{10}^2) &\leq& E_+ \leq
\frac{m_{11}^2-m_{10}^2-2 m_{11} E_-}{2(m_{11}-2 E_-)} \nonumber \\
0 &\leq& E_- \leq \frac{1}{2 m_{11}}(m_{11}^2-m_{10}^2)
\end{eqnarray}
or alternatively
\begin{eqnarray}
-[(m_{11}-E_T)^2-m_{10}^2]^{1/2} & \leq & E_\Delta \leq
[(m_{11}-E_T)^2-m_{10}^2]^{1/2} \nonumber \\
\frac{1}{2 m_{11}}(m_{11}^2-m_{10}^2) & \leq & E_T \leq m_{11}-m_{10} 
\,\,\, ,
\end{eqnarray}
where $E_T=E_++E_-$ and $E_\Delta = E_+-E_-$. (For the latter variable set, 
one must not forget to include a Jacobian factor of $1/2$.) By integrating 
over $E_\Delta$ alone, we also obtain the sum energy spectrum of the
outgoing fermion-antifermion pair.  

Before discussing our numerical results, there is an important subtlety in the
analysis related to possible degeneracies in the KK spectrum.  Notice, for 
example, that the $(1,0)$ and $(0,1)$ modes are exactly degenerate in the 
limit that $R_5=R_6$.  Moreover, these modes have identical couplings to 
fermions.  Thus, there is no way experimentally to determine 
whether an observed decay of the $(1,1)$ mode to an on-shell KK mode with 
mass $1/R$ is the decay considered above, or the analogous decay with a 
$(0,1)$ mode in the final state. More concretely, the process considered 
above might be extracted experimentally through the decay $\gamma^{(1,1)} 
\rightarrow f {\bar f} f {\bar f}$, by searching for an invariant 
mass peak in one of the $f {\bar f}$ pairs; in the case of degenerate KK 
states, there is no way kinematically to isolate the desired mode.
Since the experimental final states are identical, the amplitudes
for the $(1,0)$ and $(0,1)$ states must be added.
It is straightforward to show, however, that the sum of the nonstandard 
decay amplitudes then cancel {\em exactly} in the stated limit, and all 
hints of noncommutativity vanish! This conclusion holds in any 
Feynman diagram in which two lines of the noncommutative vertex in 
Fig.~\ref{feynman} are connected to fermions at the $\vec{y}=\vec{0}$
fixed point..   

This observation does not necessarily imply that tree-level noncommutativity 
is unobservable, but rather that one must be careful not to make too many 
simplifying assumptions on the parameters of the compactification.  If $R_5$ 
and $R_6$ differ by order one factors, as one would expect generically, then 
the degeneracy between the $(0,1)$ and $(1,0)$ states is lifted.  More 
generally, taking the parameter $\xi$ to differ from unity splits the 
degeneracy between the KK modes $(0,n)$ and $(n,0)$, while varying the 
shift angle $\phi$ away from $\pi/2$ eliminates degeneracies between the 
states $(m,n)$ and $(m,-n)$.  We will typically choose the values $\xi=0.8$ 
and $\phi=1.5$~radians for our numerical analysis so that the decay of 
interest proceeds without any ambiguity in defining the asymptotic states.
We aim to draw qualitative conclusions rather than to do a complete survey
of the available parameter space. 
\begin{figure}
  \begin{centering}
    \hfil\hspace{-10em} \epsfxsize 3.5in \epsfbox{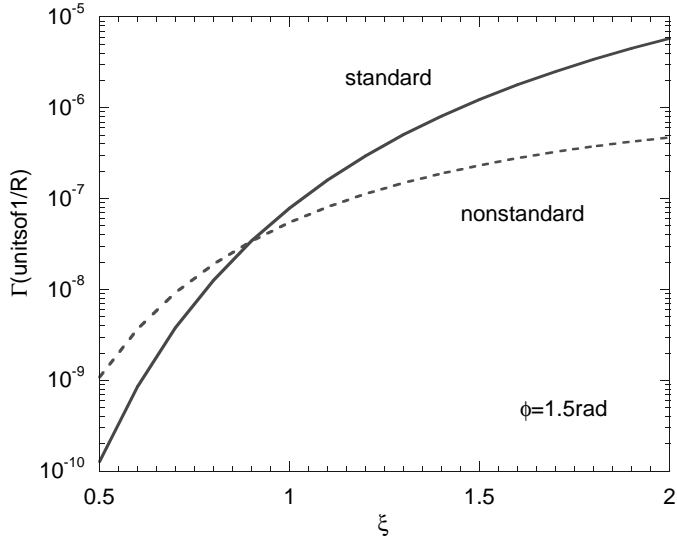} \hfill
    \caption{Contributions to the partial decay width as a function of $\xi$,
for $\sin^2(\xi_0 \theta/2 R^2)=1$ and $\phi=1.5$. }
  \label{fig:xi}
  \end{centering}
\end{figure}
Fig.~\ref{fig:xi} shows contributions to the partial decay width in units of 
the compactification scale $1/R$ as a function of $\xi$.  Since $\theta/R^2$
in Eq.~(\ref{eq:nsdecay}) is unknown, we make the optimistic choice
$\sin^2(\xi_0 \theta/2 R^2)=1$ in presenting our results. Fig.~\ref{fig:xi}
assumes there is a single decay fermion with integral charge. For 
$\xi \mbox{\raisebox{-1.0ex} {$\stackrel{\textstyle ~<~}
{\textstyle \sim}$}}0.9$, the 
nonstandard contribution to the partial width is dominant, and we expect at
least an order 100\% correction to the event rate.  On the other hand, the
branching fraction to the decay of interest is quite small. The partial 
width for the dominant two-body decay to a single fermion of integral charge, 
$\gamma^{(\vec{n})} \rightarrow f  \bar f$, is $2 \alpha m_{11}/3$,  which
is a factor of $2 \times 10^5$ larger for $\xi=0.8$ and $\phi=1.5$.  It 
follows that the branching fraction to the desired three-body decay, summed 
over fermions $f$, is $0.5\times 10^{-5}$. (We can couple to fermions 
with fractional electric charge~\cite{HAY} since matter is confined to 
a 4D subspace where the effects of noncommutativity are absent.)

\begin{figure}[ht]
  \begin{centering}
    \hfil\hspace{-10em} \epsfxsize 3.5in   \epsfbox{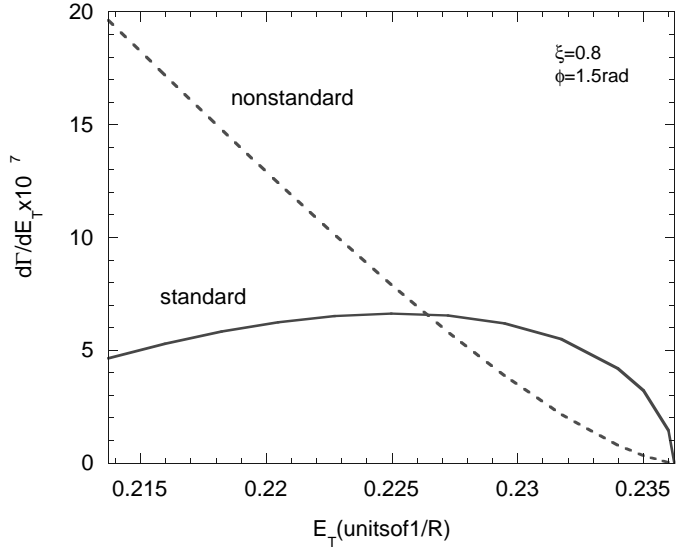} \hfill
    \caption{Contributions to the sum energy spectrum of the outgoing 
    fermion-antifermion pair.}
  \label{fig:et}
  \end{centering}
\end{figure}

Typical event rates for the s-channel production of a single KK 
mode at a large hadron collider are given in Table~\ref{tab:events}.  To 
obtain these estimates we work in the narrow width approximation and use 
CTEQ5L structure functions~\cite{5L}. For a $2$~TeV initial state produced 
at a stage 2 VLHC with 200~TeV center of mass energy and 100 fb$^{-1}$ of 
integrated luminosity we would expect to observe $49.5$ three-body decays
given the production rate shown in Table~\ref{tab:events}, 
compared to an expectation of $19.8\pm 4.4$, a $6.8$-sigma effect.  More 
importantly, the shape of the sum-energy ($E_T$) spectrum for the outgoing 
fermion-antifermion pair is dramatically different 
in the case where such an excess is not statistical, as we show in 
Fig.~\ref{fig:et}.  If collider parameters allow for a sufficient event rate,  
then an event excess particularly at smaller values of $E_T$ would be a 
telltale indication of new noncommutative interactions.
\begin{table}[ht]
\begin{tabular}{lllll}
$m_{11}$ (TeV)\qquad\qquad & $\int {\cal L}$(fb$^{-1}$)\qquad\qquad& 
$\sqrt{s}$ (TeV)
\qquad \qquad & events ($p {\bar p}$) \qquad \qquad & events ($pp$) \\ 
\hline \hline
2       & 100 fb$^{-1}$  &  14 & $3.8 \times 10^5$  & $9.3 \times 10^4$ \\
$\cdot$ &   $\cdot$      &  80 & $3.2 \times 10^6$  & $2.7 \times 10^6$ \\
$\cdot$ &   $\cdot$      & 140 & $6.5 \times 10^6$  & $6.1 \times 10^6$ \\
$\cdot$ &   $\cdot$      & 200 & $1.0 \times 10^7$  & $9.9 \times 10^6$ \\
4       &   $\cdot$      &  14 & $1.7 \times 10^4$  & $1.3 \times 10^3$ \\
$\cdot$ &   $\cdot$      &  80 & $3.4 \times 10^5$  & $2.2 \times 10^5$ \\
$\cdot$ &   $\cdot$      & 140 & $6.7 \times 10^5$  & $5.6 \times 10^5$ \\
$\cdot$ &   $\cdot$      & 200 & $1.1 \times 10^6$  & $9.6 \times 10^5$ \\
\hline\hline 
\end{tabular}
\caption{Number of KK modes produced at very large hadron colliders.}
\label{tab:events}
\end{table}      

Whether or not a future high energy collider would provide for a
sufficient event rate to detect noncommutativity in the 
$\gamma^{(\vec{n})} f {\bar f}$ decay channel is a difficult
question.  The previous example suggests an affirmative answer, though
the value of $m_{11}=2$~TeV is slightly low compared to most precision
electroweak bounds on the compactification scale~\cite{pew}. (The mass
spectrum is compatible with current direct collider searches.)  Indirect
bounds rely strongly on the assumption that no other physics contributes
to the relevant low-energy (usually $Z$-pole) observables.  It is likely
that these bounds could easily be relaxed by $O(1)$ factors if Planck-
suppressed, higher-dimension operators are taken into account, given
that the Planck scale is low. If one is more conservative and considers
the safer $4$~TeV results in Table~\ref{tab:events}, then the event rate
drops at a $200$~TeV VLHC by about an order of magnitude ($4.8$ events
with an expectation of $1.9\pm 1.4$). It is likely in this case that a
higher integrated luminosity would be required before any definite
conclusions could be reached.

To summarize, we have found that noncommutativity in the decays
$\gamma^{(\vec{n}_a)} \rightarrow \gamma^{(\vec{n}_b)} \, f \, \bar{f}$
is potentially observable at a very large hadron collider, providing one
is lucky with the parameters of the theory.  In the next section, we will
see that an easier signal is an enhancement in the pair
production of some of the lighter KK modes.


\section{Pair Production}


Pair production of KK photons at colliders can occur through standard
diagrams involving no noncommutative interactions.  Additionally, pair
production can proceed through the triple 6D-photon
vertex which does not exist without noncommutativity.  

\begin{figure}[ht]
\centerline {\epsfxsize 3.3 in \epsfbox{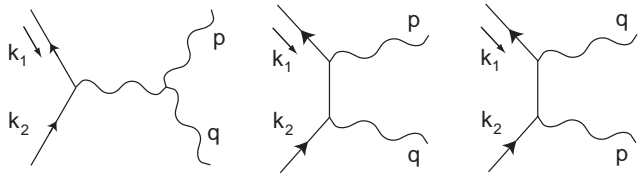}   }
\caption{Feynman diagrams for production of KK pairs, including the
noncommutative triple photon vertex.}
\label{pair}
\end{figure}

The standard  and non-standard pair production processes are shown at
the parton level in Fig.~\ref{pair}.  Their matrix elements may be
obtained from those of the decay process studied in the last section
through crossing symmetry.

We present the cross section again for the case that the Kaluza-Klein
states are the $\vec n = (1,0)$ and $(1,1)$ modes.  While noncommutativity
also affects the production of $(0,1)$-$(1,0)$ pairs, the nonstandard
diagram in this case involves contributions from different intermediate
states that tend to cancel, suppressing the rate.  At the parton level, 
for the final states we have chosen, the nonstandard cross section is
\begin{eqnarray}
\sigma_{NS} (\hat s) &=& {\pi\alpha^2 \over 3 \hat s^2 } \lambda
    \left[ (m_{21}^2 - m_{01}^2)/(m_{11}m_{10}) \over 
                  (\hat{s}-m_{21}^2)(\hat{s}-m_{01}^2) \right]^2
            \sin^2 \left( \xi_0\theta\over 2R^2 \right) 
                                  \nonumber   \\    
                              \times
\Big\{ &\hat s^4& +8(m_{11}^2+m_{10}^2) \hat s^3 
      -[18m_{11}^4+32m_{11}^2 m_{10}^2+18 m_{10}^4] \hat s^2
                                      \nonumber \\    
                              &\quad&
+ 8 (m_{11}^6-4m_{11}^4 m_{10}^2-4m_{11}^2 m_{10}^4+m_{10}^6) \hat s 
                                      \nonumber \\ 
                              &\quad&
+ ( m_{11}^2-m_{10}^2)^2 (m_{11}^4 +10m_{11}^2 m_{10}^2 +m_{10}^4)
\Big\}     \ ,
\end{eqnarray}

\noindent where $\hat s$ is the partonic center-of-mass (CM) energy
squared and $\lambda$ is defined in terms of the CM 3-momentum of
either final state particle, $|\vec p| = \lambda/2\sqrt{\hat s}$ with
\begin{equation}
\lambda = \sqrt{\hat s^2 - 2 \hat s (m_{11}^2+m_{10}^2) + 
(m_{11}^2-m_{10}^2)^2} \ .
\end{equation}

\noindent The
initial partons are treated as massless.  The parton level cross section
for the standard  process is  
\begin{eqnarray}
\sigma_{S}(\hat s) &=& {16\pi\alpha^2\over \hat s^2} 
   \Bigg\{ -\lambda {m_{11}+m_{10} \over m_{11}} 
                                                \\    
                              &\quad&
+ {\hat s^2 +\left( m_{11}^2+m_{10}^2 \right)^2 \over \hat s - m_{11}^2
-m_{10}^2 }
  \ln{ \hat s-m_{11}^2-m_{10}^2 + \lambda \over \hat s-m_{11}^2-m_{10}^2 
- \lambda }  
                                                        \Bigg\}  \ .
                                                          \nonumber
\end{eqnarray}

The collider cross section is
\begin{eqnarray}
&\sigma&(s,AB \rightarrow \gamma_{11} \gamma_{10} X)
  = \int_\tau^1 dx_1 \int_{\tau/x_1}^1 dx_2
                                      \nonumber \\    
                              &\times&
\frac{1}{3} \sum_q \left[ f_{q/A}(x_1) f_{\bar q/B}(x_2)
                + f_{\bar q/A}(x_1) f_{q/B}(x_2) \right]
                                      \nonumber \\    
                              &\times&   
\left\{ e_q^4 \sigma_{S}(\hat s) + e_q^2 \sigma_{NS}(\hat s) \right\}
\ ,
\end{eqnarray}

\noindent where $\hat s = x_1 x_2 s$, $\tau = \hat s_{min}/s$, and 
$\hat s_{min}$ is the square of the sum of the KK excitation
masses, or
\begin{equation}
\hat s_{min} = \left(  m_{10}+m_{11}  \right)^2  \ .
\end{equation}

\noindent Also, $f_{q/A}(x) = f_{q/A}(x,\mu)$ are the parton distribution
functions for quark $q$ in hadron $A$ evaluated at renormalization scale
$\mu$, and the $1/3$ is from color averaging.

\begin{figure}[t]
\centerline {\epsfxsize 3.5 in \epsfbox{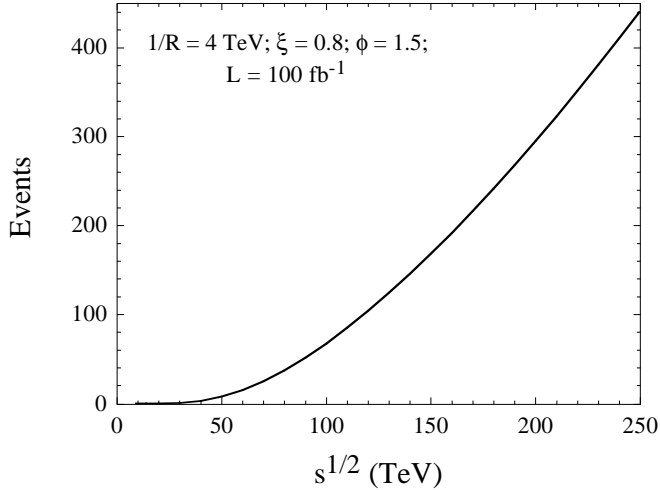}   }
\caption{Event rate at a proton-proton collider for the production of 
the KK pair $\gamma_{11}+\gamma_{10}$ vs. the $pp$ center-of-mass energy
$\sqrt{s}$.}
\label{fig:pevents}
\end{figure}

We evaluated the cross section for a proton-proton collider again using
the CTEQ5L parton distribution functions at a fixed scale $\mu = 2$
TeV, with $\xi = 0.8$, $\phi=1.5$ and $\sin^2(\xi_0 \theta/2 R^2)=1$  . 
Fig.~\ref{fig:pevents} shows the event rate 
over a range of center of mass energies for $1/R=4$~TeV and $100$~fb$^{-1}$ of
integrated luminosity.  At a $200$~TeV VLHC, for example, we find
$294$ events where there is an expectation without noncommutativity of
$165 \pm 12.8$, a $10.1$ sigma effect.  This is a significant signal 
for a choice of $1/R$ that is consistent with current indirect bounds
on the compactification scale.  As in the example of the decays, what
is more significant is the way in which noncommutativity emerges.  For
example, production of $(1,0)$-$(1,0)$ pairs receives {\em no}
noncommutative corrections while production of $(1,0)$-$(1,1)$ pairs
does.  Comparison of these channels may help eliminate uncertainty
originating, for example, from parton distribution functions.

\section{Conclusions}

We have explored the phenomenological consequences of 
noncommutativity in extra spatial dimensions.  By restricting
noncommutativity to the bulk, we avoid conflict with the stringent 
experimental limits on the violation of four-dimensional Lorentz invariance, 
which otherwise force the magnitude of noncommutativity to be small.  We 
constructed an explicit example, based on the orbifold $T^2/ \mathbb Z_2$, 
to illustrate the effects of spatial noncommutativity in 6D QED with fermions 
confined to an orbifold fixed point.  Notably, we find new three- and 
four-point couplings involving KK excitations of the photon, exclusively.  
Since all fields involved in these interactions have TeV-scale masses, we 
confirm that collider signals are most promising at a VLHC, rather than at the
LHC.  In particular, we find order 100\% corrections to the three-body
decays $\gamma^{(\vec{n})}\rightarrow  \gamma^{(\vec{m})} f {\bar f}$
and to the pair production $f {\bar f}\rightarrow 
\gamma^{(\vec{m})}\gamma^{(\vec{n})}$, with $\vec{m} \neq \vec{n}$. The
former might be discernible at a VLHC, if one is lucky with model
parameters, and yield a strikingly different $f {\bar f}$ sum energy
spectrum then if the noncommutative interactions are absent.  
Pair production has a better chance of yielding a statistically
significant excess for TeV-scale KK masses.  Observing order 100\% corrections
to the production of certain pairs of KK modes at a VLHC while finding no 
corrections to others would provide a clear signal of noncommutativity 
in the bulk.

%
\begin{acknowledgments}
We thank the National Science Foundation for support  under Grant 
No.\ PHY-9900657.  In addition, C.D.C thanks the Jeffress Memorial 
Trust for support under Grant No.~J-532.
\end{acknowledgments}


\end{document}